\documentstyle[12pt,twoside,fleqn,espcrc1,psfig]{article}

\title{The rp-process and new measurements of $\beta$-delayed proton decay
of light Ag and Cd isotopes}

\author{
G. Raimann\address{Department of Physics, The Ohio State University,
Columbus, OH 43210, USA}, 
M.J. Balbes$^{\rm a}$\thanks{Present address: Computerized Medical Systems,
St.\ Louis, MO, USA},		
R.N. Boyd$^{\rm a}$\address{Department of Astronomy, The Ohio State University,
Columbus, OH 43210, USA},
D. Cano-Ott\address{Instituto de F\'\i{}sica Corpuscular, 46100 Burjassot
Valencia, Spain},
R. Collatz\address{Gesellschaft f\"ur Schwerionenforschung, D-64291
Darmstadt, Germany}, 
A. Guglielmetti$^{\rm d}$\thanks{Present address: Istituto di Fisica, 
I-20133 Milano, Italy},
M. Hellstr\"om$^{\rm d}$\thanks{Supported by the Alexander von Humboldt
Foundation}, 
M. Hencheck\address{School of Math \& Science, Chadron State College,
Chadron, NE 69337, USA}, 
Z. Hu$^{\rm d}$,
Z. Janas\address{Department of Physics, University of Warsaw, PL-00681
Warsaw, Poland}, 
M. Karny$^{\rm f}$, 
R. Kirchner$^{\rm d}$,
J. Morford$^{\rm e}$, 
D.J. Morrissey\address{Department of Chemistry, Michigan State University,
East Lansing, MI 48824, USA},
E. Roeckl$^{\rm d}$, 
K. Schmidt$^{\rm d}$,
J. Szerypo$^{\rm f}$, 
A. Weber$^{\rm d}$
}

\begin{document}
\maketitle

\begin{abstract}

Recent network calculations suggest that a high temperature rp-process could
explain the abundances of light Mo and Ru isotopes, which have long challenged
models of p-process nuclide production. Important ingredients to network
calculations involving unstable nuclei near and at the proton drip line are
$\beta$-halflives and decay modes, i.e., whether or not $\beta$-delayed proton
decay takes place. Of particular importance to these network calculation are
the proton-rich isotopes $^{96}$Ag, $^{98}$Ag, $^{96}$Cd and $^{98}$Cd. We
report on recent measurements of $\beta$-delayed proton branching ratios for
$^{96}$Ag, $^{98}$Ag, and $^{98}$Cd at the on-line mass separator at GSI.

\end{abstract}

\section{THE QUEST FOR THE P-NUCLEI}

The p-process elements, proton-rich nuclei heavier than iron that cannot be
produced by neutron-capture (r- and s-processes), are among the least abundant
elements in the cosmos. Their synthesis has been modelled as an rp-process
\cite{Bur57}, a series of (p,$\gamma$) reactions on existing r- and s-process
seeds in a hot hydrogen rich environment, or a $\gamma$-process \cite{Woo78},
which creates proton-rich nuclei by photodissociation reactions, ($\gamma$,p),
($\gamma$,n) and ($\gamma,\alpha$), thought to take place in type II
supernovae. This $\gamma$-process requires temperatures of the order of $T_9$ =
2 -- 3, starts from r- and s-process seed nuclei and runs on timescales of
order a second. It appears to explain the abundance of heavy p-nuclei ($A >
108$) quite well (for shortcomings of the $\gamma$-process and further
references, see e.g.\ \cite{Hof96}), but always severely underproduces the
lighter p-nuclei ($A \le 108$) by factors of the order of 10 to 100. 

Some of the lighter p-nuclei, specifically $^{92}$Mo, $^{94}$Mo, $^{96}$Ru and
$^{98}$Ru, are much more abundant than their heavier (and lighter)
counterparts. If these are produced by the rp-process (see
\cite{Aud75,Can93,Wor94}; recent reviews given by \cite{Lam92} and
\cite{Mey94}), high temperatures (of the order $T_9 = 1$) and hydrogen richness
($\rho Y_p$ of the order $10^3$ to $10^4$ g cm$^{-3}$) are required. 

Accreting neutron stars or white dwarfs in binary systems and Thorne-\.Zytkow
objects have been suggested as possible sites for the rp-process
\cite{Can93,Bie91}. In the neutron star scenario, hydrogen rich material flows
from the partner (usually a red giant) to the neutron star and reaches high
enough temperatures during the accretion process. In a Thorne-\.Zytkow object,
a neutron star is merged with a red giant. There will be a strongly convective
hydrogen-rich environment at high temperatures directly above the neutron star.
A recent study \cite{Hen96} assuming a pulsed rp-process appears to be capable
of producing substantial amounts of light p-nuclei, especially the important
and difficult to create Mo and Ru isotopes. This study provides the incentive
to determine experimentally the decay modes of some of the progenitors of
$^{96}$Ru and $^{98}$Ru, namely $^{96}$Ag, $^{98}$Ag and $^{98}$Cd. 

Another recent investigation \cite{Hof96} found that, in a new kind of
p-process in the neutrino-driven wind following the delayed explosion of a
supernova type II, some light p-nuclei may actually be coproduced with
r-process isotopes. This process would take place in addition to the usual
$\gamma$-process, which would account for the heavier p-nuclei. However, little
$^{94}$Mo and $^{96,98}$Ru are synthesized in this process.

\section{THE EXPERIMENT}

The abundances of the Mo and Ru isotopes produced in network calculations
depend sensitively on the decay modes of the progenitors. While $^{96}$Cd has
not yet been identified, only half-lives and $\gamma$-ray properties have been
established for $^{96}$Ag \cite{Kur82,Sch95}, $^{98}$Ag \cite{Kur82} and
$^{98}$Cd \cite{Plo92}. In this experiment, we have re-investigated the decay
properties of these nuclides, including $\beta$-delayed proton decays. A
$^{58}$Ni beam ($5.9 - 6.7$ MeV/u, $\approx 40$ particle nA) from the GSI
UNILAC accelerator interacted with a production target of $^{50}$Cr (for $A =
96, 98$) and $^{58}$Ni (for an $A = 114$ calibration run) in a FEBIAD-B2-C ion
source \cite{Kir87}. The reaction products were mass analyzed using the GSI
on-line mass separator. 

The unique design of the mass separator allowed three setups to be used in
parallel (with different masses selected). Setup (a) allowed implantation of
the selected isotope onto a tape, which was then transported into a measuring
station with a HP-Ge $\gamma$-detector and a $\Delta E$-$E$ silicon detector
telescope for protons. Setup (b) consisted of two identical silicon detector
telescopes. The beam could be switched between them, and beam implantation into
thin foils in close geometry allowed measurement of growth and decay of proton
activity. Setup (c) featured a $\beta$-telescope made of two plastic
scintillators. The beam was implanted onto a tape directly in front of the
measuring station. 

With these setups, the $\beta$-delayed proton decay rates, $b_{\beta p}$, of a
given nucleus can be obtained in the following four ways: (i) both proton and
$\gamma$-rates obtained at setup (a); (ii) proton rate from (b), and
mass-separated beam intensity from $\gamma$-measurement at (a); (iii) proton
rate from (b), and mass-separated beam intensity from $\beta$-measurement at
(c); (iv) proton rate from (b), and mass-separated beam intensity from (c). 

In addition to providing a continuous beam, the FEBIAD-B2-C ion source can be
operated in bunched mode \cite{Kir87}. Some ions with sufficiently long
half-lives, e.g.\ $^{98}$Ag, can be collected in a cold pocket, thus
suppressing Ag relative to Cd by a factor of $\sim$ 10--20 (``Cd anti-bunch'').
Subsequent heating of the cold spot can release substantial quantities of the
stored ion species, enhancing Ag over Cd by a factor of $\sim 100$ (``Ag
bunch'').

\section{RESULTS}

A mass 114 beam was used to calibrate each setup independently, based on the
known decay of $^{114}$Cs \cite{Roe80}. The four methods described above gave
consistent $b_{\beta p}$-values, and the weighted average was in good agreement
with literature (Table 1). 

\begin{table}[htb]
\vspace*{-5mm}
\caption{Production rates and $\beta$-delayed proton branching ratios.}
\label{t1}
\begin{tabular*}{\textwidth}{@{}l@{\extracolsep{\fill}}cccccc}
\hline
Isotope & \multicolumn{2}{c}{Rates (s $\cdot$ 10 particle nA)$^{-1}$} &
          \multicolumn{2}{c}{$\beta$-delayed proton branching ratio (\%)} \\
	   & Beam			& Protons 	
					& This work	
					& Literature	\\
\hline
$^{96}$Ag  & 2.9(8) $^{\rm a}$		& 0.11(2) $^{\rm c}$ 
					& 3.7(9)  $^{\rm e}$ 
					& 8.0(23) \cite{Kur82} \\
	   &				&
					&		
					& $11.9(26)$ \cite{Sch95} \\
$^{98}$Ag  & 59(8) $^{\rm a}$		& $(6.4^{+3.5}_{-2.7})\times 10^{-4}$ 	
						$^{\rm d}$ 
					& $(1.1^{+0.5}_{-0.4})\times 10^{-3}$ 
						$^{\rm f}$ \\
$^{98}$Cd  & $>1.2$ $^{\rm b}$		& $< 3.05 \times 10^{-4}$ $^{\rm d}$ 
					& $< 2.5  \times 10^{-2}$ $^{\rm g}$ \\
$^{114}$Cs & 11.8(16) $^{\rm a}$	& 1.1(2) $^{\rm c}$ 
					& $8.7(13)$ $^{\rm e}$
					& $7(2)$ \cite{Roe80} \\
\hline
\end{tabular*}
{\small
$^{\rm a}$ Weighted average from setups (a) and (c). 
$^{\rm b}$ From setup (a). 
$^{\rm c}$ Weighted average from setups (a) and (b). 
$^{\rm d}$ From setup (b). 
$^{\rm e}$ Weighted average from methods (i) to (iv). 
$^{\rm f}$ Weighted average from methods (ii) and (iii). 
$^{\rm g}$ Method (ii). 
}
\vspace*{-10mm}
\end{table}

\begin{figure}[hbt]
\begin{minipage}[t]{80mm}
\vspace*{-0.5cm}
\psfig{file=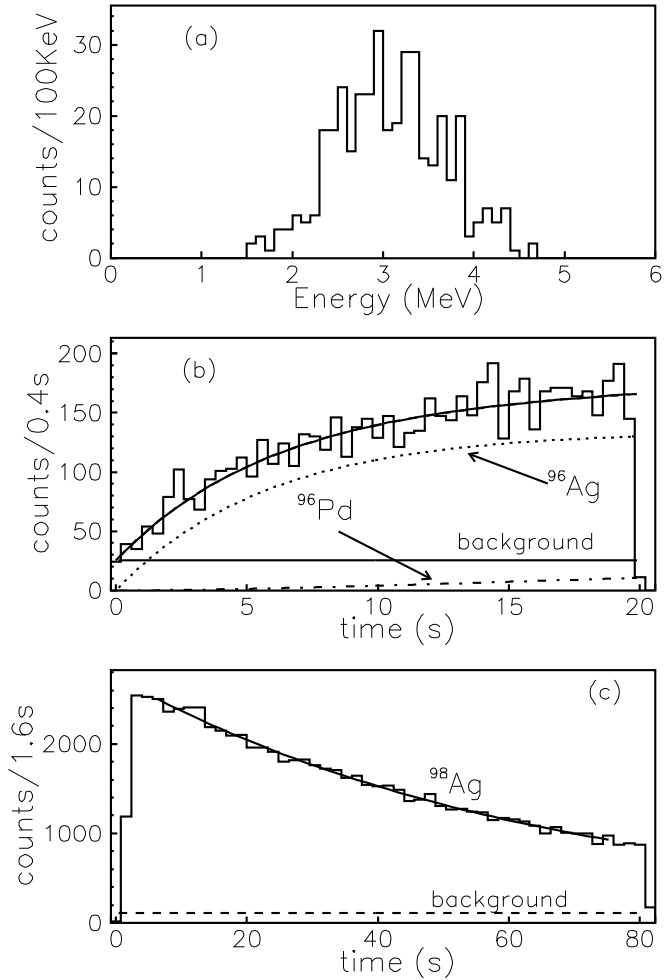,width=100mm}
\end{minipage}
\hspace{\fill}
\begin{minipage}[t]{55mm}
\caption{(a) $^{96}$Ag $\beta$-delayed proton energy spectrum.
(b) Time distribution of $\beta$-particles from $^{96}$Ag. The fit includes
$^{96}$Ag, $^{96}$Pd (daughter activity) and background. 
(c) Time distribution of $\beta$-activity from $^{98}$Ag. The fit includes
$^{98}$Ag and background. }
\label{fig1}
\end{minipage}
\end{figure}

For the mass 96 measurements, the ion source was operated in continuous beam 
mode. The observed protons were assigned to the decay of $^{96}$Ag, since
its production cross section is predicted to be much larger than that of
$^{96}$Cd. Also, odd-odd nuclei such as $^{96}$Ag are expected to have larger
$b_{\beta p}$-values than neighboring even-even nuclei such as $^{96}$Cd due to
the combined action of larger decay-energy window and enhanced $\beta$-feeding
of higher-lying (4qp) states. See Table 1 and Fig.\ 1. 

In the measurement of mass 98, Ag was bunched. The release time window was
only 1.5 to 2.0 seconds over a total cycle period of 81 to 85 s. A total of
8 protons was measured in setup (b) over a period of about 8.5 hours. The
corresponding Cd anti-bunch measurement resulted in no protons. All 8 protons
were therefore assigned to $^{98}$Ag. For $^{98}$Cd Table 1 reports a
1 $\sigma$ upper limit. 

The apparent discrepancy of the new $b_{\beta p}$-value for $^{96}$Ag with
literature may be due to the fact that different production reactions
($^{58}$Ni + $^{50}$Cr in this experiment, $^{40}$Ca + $^{60}$Ni in
\cite{Kur82,Sch95}) may result in a different population of the ground state
and isomeric state of $^{96}$Ag, both of which may undergo $\beta$-delayed
proton decay. 

In conclusion, the present experiment shows that $\beta$-delayed
proton branching ratios of $^{96}$Ag, $^{98}$Ag and $^{98}$Cd are small,
and thus the abundances of $^{96}$Ru and $^{98}$Ru are directly related to
the abundances of their progenitors in rp-process production of Ru. 

Supported in part by NSF grant PHY 95-13893 and PHY 95-28844. The polish
authors acknowledge partial support by grant KBN 2 P302 06.

\end{document}